\documentclass{epl}
\date{\today}
\shorttitle{Glassy phase of RNA secondary struct.}
\title{Nature of the glassy phase of RNA secondary structure}
\author{F. Krzakala
\and M.\ M\'ezard
\and M.\ M\"uller}
\institute{
Laboratoire de Physique Th\'eorique et Mod\`eles Statistiques,\\
Universit\'e Paris-Sud, b\^atiment 100, F-91405 Orsay,
France.
}

\begin{document}

\maketitle

\begin{abstract}
	We characterize the low temperature phase of
a simple model for RNA 
secondary structures by determining the typical energy scale $E(l)$ of 
excitations involving $l$ bases. 
At low enough temperatures, including $T=0$,
we find  a scaling law $E(l)\sim l^\theta$ with a small exponent $\theta$.
Above a critical temperature, there is a different
phase characterized by a relatively flat free energy landscape resembling that of a homopolymer with a scaling exponent $\theta=1$. These results strengthen the evidence in favour of the existence
of a glass phase at low temperatures. 
\end{abstract}

\section{Introduction}  
  The folding of RNA or single stranded DNA as described by its secondary 
structure is both a relevant problem in molecular biology and a 
challenging task for the statistical mechanics of disordered systems. Several 
authors~\cite{Higgs96,PagnaniParisi00,Hartmann01,PagnaniParisi00Reply,BundschuhHwa01a,BundschuhHwa01b} 
have recently addressed the topic and put forward some evidence for the 
existence of a glassy phase at low temperatures.
Numerical studies of the specific heat demonstrate the existence of a higher order phase transition
\cite{PagnaniParisi00}. The nature 
and properties of the low temperature phase are less clear because of large finite size corrections. While the
overlap distribution is certainly broad for systems of up to 1000 bases~\cite{Higgs96,PagnaniParisi00}, indicating a kind of glassy phase, its asymptotic behaviour for long sequences cannot be deduced reliably from
the present simulations\cite{Higgs96,PagnaniParisi00,Hartmann01,PagnaniParisi00Reply}. Bundschuh and Hwa~\cite{BundschuhHwa01a,BundschuhHwa01b} have  
recently argued in favour of the existence of a glass phase. They 
showed analytically that in the disordered case the asymptotic 
pre-exponential scaling of the partition function cannot be the 
same as for homogeneous RNA at low temperatures. They also showed 
that the system of two attractively coupled replicas of the same 
disordered sequence exhibits a phase transition from a strongly 
coupled low temperature phase to a phase at high temperatures 
where the replicas are essentially independent. Both results favour the existence of a glass transition at finite temperature, 
but a rigorous proof is still missing. Numerically, the same authors 
characterize the RNA conformation via the free energy cost of an 
imposed pairing (pinching) of two bases. Concentrating on the largest 
possible pinching excitations in the groundstate of a given RNA sequence 
they argue in favour of an excitation energy scale that grows 
logarithmically with the number $N$ of bases in the sequence, 
a weak power law not being ruled out. However, it is not clear in 
what sense the excitations created by a single pinch are typical.

	In this paper, we propose a different numerical method to 
study the scale dependence of excitations as was originally 
introduced in the context of spin 
glass models~\cite{CaraccioloParisi90a,Mezard90,PalassiniYoung00a}. The aim is 
to determine the typical free energy scale $E(l)$ of excitations involving 
the bonds of $l$ bases. Rather than changing boundary conditions 
via pinching, we introduce a perturbation in the bulk which 
allows for a better control of the actual size of excitations.

	We find that the low temperature regime is governed by 
excitations which scale like $E(l)\sim l^\theta$ where 
$\theta\approx 0.23$ while the high temperature phase behaves like a homopolymer with scaling exponent $\theta=1$. The  
change of $\theta$ indicates a 
phase transition between a liquid-like high temperature phase and a strongly correlated glass-like phase 
at low temperature. In the former, excitations consist of independent rearrangements
of individual bases, and the paired pieces of the polymer can 
slide along each other, while in the latter the system is locked in a favourable 
secondary structure, and low-lying excitations involve correlated 
rearrangements of bonds. Note that our description is restricted to the thermodynamic and static properties of RNA. In particular, within our approach we cannot deal with the dynamics of RNA which is expected to be very slow in the low temperature phase~\cite{IsambertSiggia00}. 

\section{The model}
The RNA-strand is characterized by its (quenched) base sequence, real RNA being composed of the four constituents A,C,G and U. The single stranded polymer will fold back onto itself to form local double helices of stacked base pairs as found in double stranded DNA. The pattern of base pairings is known as the secondary structure of RNA. The Watson-Crick base pairs A-U and C-G have the strongest tendency to bind, and in a first approximation 
one can neglect other pairings. In this paper we concentrate on the secondary structure without taking into account the three-dimensional ternary structure of the polymer whose typical energy scale is significantly lower than that of the base pairing~\cite{ZukerSankoff84,BundschuhHwa01b}.

The glass phase which we will describe exists at the level of the secondary structure. One might imagine another glass phase appearing at the level of the ternary
structure, in the form of the possible existence of many metastable spatial
arrangements of the molecule, while the secondary structure would be unique.
 We do not address this question here.

Only two restrictions from the spatial structure  are retained: The rigidity of the polymer chain is taken into account by requiring a minimal distance $|j-i|\geq s$ between two bases forming a bond to avoid very small loops of the linking strand.
Furthermore, pairings of bases $\{i,j\}$ and $\{k,l\}$ with $i<k<j<l$ are known to be very rare in real RNA for topological reasons.
We exclude such pseudoknots altogether in this paper as they can be regarded as a small perturbation. 
They could in principle be taken into account systematically, 
using the approach of  ~\cite{OrlandZee01}.

To simplify the problem even further we consider 
a variant of
the model defined in~\cite{PagnaniParisi00} with sequences $(b_i)_{i=1,\dots, N}$ of only two different
 species, $b_i = A$ or $B$, with bond energies $e(A,A)=e(B,B)=-1$ and $e(A,B)=e(B,A)=-2$, which is expected to capture the essential physics of RNA. 
To avoid an artifact of the two letters model for $s=1$ (see \cite{BundschuhHwa01b}), we choose $s=2$ as in earlier work~\cite{PagnaniParisi00}.
A natural interpretation of
this model is that  $A$ and $B$ correspond to small subsequences of a real RNA-strand and the energies describe their effective pairing affinities. To mimic the fluctuations of the latter we add a noise to all bond energies, $e_{i,j}\rightarrow e_{i,j}+\eta_{i,j}$ where $e_{i,j}=e(b_i,b_j)$ and $\eta_{i,j}$ is a uniformly distributed variable in $[-1/2,1/2]$.
This lifts the exact degeneracy and 
ensures that the ground state is unique, which
is certainly true in real RNA (or 
when using more realistic
rules for secondary structures energies), and is also useful technically.
In parallel, we studied models with different base pairing energies, and with couplings $e_{i,j}$ that are independently drawn from a given continuous distribution without reference to a base sequence. The results were qualitatively the same in all models.
   
Taking the energy of a secondary structure ${\mathcal S}$ to be the sum of pairing energies, 
$H({\mathcal S})=\sum_{\{i,j\}\in {\mathcal S}}e_{i,j}$,
the partition 
function of an RNA strand with $N$ bases is given by
\begin{equation}
	Z_N=\sum_{{\mathcal S}_N}\exp [-\beta H({\mathcal S}_N)],
\label{Zdef}
\end{equation}
where the sum extends over all permissible secondary structures ${\mathcal S}_N$. Let us represent a secondary structure ${\mathcal S}$ by the set of numbers $\{p_i\}$ ascribing to each base $i$ the base $p_i=j$ that it is paired to, or $p_i=-1$ if it is unpaired. We can then define an overlap between two secondary structures ${\mathcal S}_\alpha$ and ${\mathcal S}_\beta$ (for the same base sequence) as

\begin{equation}
	q^{\alpha\beta}=\frac{1}{N}\sum_{i=1}^{N}\delta_{p_i^\alpha p_i^\beta}
\label{overlap}
\end{equation}
which is normalized to $q^{\alpha\alpha}=1$.

Earlier investigations~\cite{Higgs96,PagnaniParisi00} on very similar models
concentrated on the probability distribution of the overlap, $P(q)$, which was found to exhibit non-selfaverageness, i.e., $P(q)$ depends on the realization of the sequence disorder. Such properties are interpreted as signatures~\cite{MarinariNaitza98b} of a glass phase since they require correlations over large parts of the system, implying a divergent correlation length in the low temperature phase. 

	However, from the numerical data it was not clear whether or 
not in the thermodynamic limit $P(q)$ tends to a single delta peak, $\delta(1-q)$, or to a nontrivial function as in mean field spin glasses. The latter would imply the existence of arbitrarily large excitations of energy $O(1)$ and thus a free energy exponent $\theta=0$. For positive $\theta$, however, the groundstate dominates at sufficiently low temperature and the overlap is peaked at $q=1$ in the thermodynamic limit. 
The system can nevertheless possess a glass phase where favourable 
free energy valleys are separated by high barriers, but differ in free
energies by terms of order $N^\theta$. This is what happens for instance 
in the case of a directed
polymer in random media (DPRM) \cite{HalpinHealyZhang95}, where $\theta=1/3$ in $1+1$ dimensions,
and we shall show below that the situation is similar in our RNA model at 
low temperatures. 
(Let us notice that, in the case of special 
sequences made up of A and G in the first half of the sequence and C and U in the second half, the RNA model can be mapped onto the
DPRM \cite{HwaLassig97}.) In such a case, the $\epsilon$-coupling method described 
below is a choice method to detect the glass phase and to compute $\theta$.

\section{$\epsilon$-coupling at $T=0$}
	Our approach allows a rather direct analysis of the low-lying 
energy landscape. The basic idea is to introduce a small perturbation 
in the bulk which repels the system from its groundstate ${\mathcal S}_0$
\cite{DrosselBokil00,PalassiniYoung00a}. 
More precisely, one considers the new Hamiltonian
\begin{equation}
	H'({\mathcal S};\epsilon)=H({\mathcal S})+\epsilon q({\mathcal S},{\mathcal S}_0)
\label{Heps}
\end{equation}
where the repulsive coupling $\epsilon>0$ will be tuned appropriately with system size. The energy of the original groundstate 
is increased by $\epsilon$ whereas every other state is shifted 
by a smaller amount. The groundstate 
${\mathcal S}_\epsilon$ of $H'({\mathcal S};\epsilon)$ is a low-lying excitation of the original system and has the largest distance $1-q({\mathcal S}_\epsilon,{\mathcal S}_0)$ from the groundstate at the given 
excitation energy.

	Let us suppose that the typical  energy scale of excitations 
involving 
the change of the pairing of $l$ bases is $E(l)\sim l^\theta$, or more precisely, 
that the disorder-averaged probability distribution of  energies obeys a 
scaling law $P[E(l)]=\frac{1}{l^\theta}f\left[\frac{E(l)}{l^\theta}\right]$. 
Under the assumption that $f(0)$ is finite and $\theta<1$ one expects the average fraction of bases involved in an excitation to scale as
\begin{equation}
	\overline{1-q({\mathcal S}_\epsilon,{\mathcal S}_0)}\sim \int_0^{\epsilon/N^\theta} f(E) dE
\label{1-q1}
\end{equation}
since large 
scale excitations of the order of the system size $N$ dominate 
the disorder average (indicated by the overbar). 
Note that $\theta<1$ implies that, for fixed $l$, excitations composed of several independent 'elementary' excitations are usually higher in energy so that the contribution of 'elementary' excitations dominates. 
As we shall see, the data fits well to the more general dependence
\begin{equation}
	\overline{1-q({\mathcal S}_\epsilon,{\mathcal S}_0)}=g(N)\Phi( \frac{\epsilon}{N^\theta}),
\label{1-q}
\end{equation}

where $g(N)$ has been introduced to account for finite size
effects and is subject to two boundary conditions: 
For $\epsilon \gg N^\theta\gg 1$, $\overline{q}$ vanishes, 
and thus $g(N\rightarrow\infty)=1/\Phi(\infty)\equiv1$. 
On the other hand, for $\epsilon \ll N^\theta$ and small $N$ 
one has to recover a behaviour linear in $\epsilon/ N^\theta$ which implies 
that $g(N\rightarrow 0)\rightarrow {\rm const.}$ and $\Phi$ behaves linearly. 

\begin{figure}
\resizebox{0.99\textwidth}{!}{
  \includegraphics{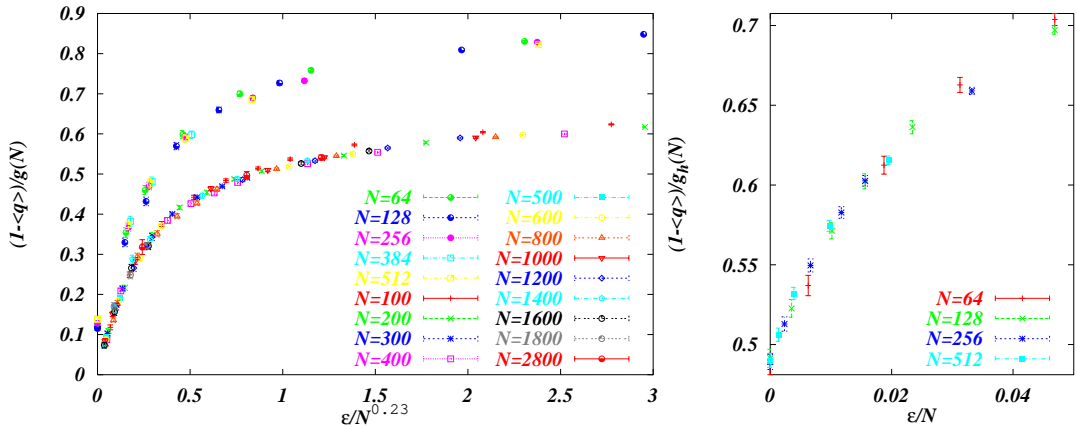}}
\caption{Left: Scaling plot for $\epsilon$-coupling at $T=0$ and $T=0.05$ (shifted vertically by 0.3) with $\theta=0.23$, giving  
 the distance between the coupled secondary structures versus the strength
of the repulsion in reduced units.
Right: Scaling plot at T=0.25 with $\theta=1$. 
}
\label{Fig_scaling}
\end{figure}

	The algorithmic implementation of $\epsilon$-coupling is rather straightforward. The ground state of a particular realization of $e_{i,j}$ can be found recursively in $O(N^3)$ time~\cite{ZukerSankoff84,Higgs96}. The new Hamiltonian (\ref{Heps}) is equivalent, up to an irrelevant constant, to a change of the bond energies and the new groundstate can be found in the same way as before. We have used $100$ to $1000$ samples for sizes ranging from $100$ to $2800$ and with $\epsilon$ ranging from $0.2$ to $10.0$. In Fig.\ref{Fig_scaling} we have recast our data into a scaling plot using the trial function $g(N)=1+c/(1+N/N^*)$ for the finite size effects. 
We found an optimal value $\theta=0.23\pm0.05$ for the energy exponent while typical values for the parameters of $g(N)$ are $c\approx 0.6$ and $N^*\approx 300-1000$. We also tried to collapse the curves with scaling variables $\epsilon/\phi(N)$ where $\phi(N)=\log(N/N^*)$ as suggested in \cite{BundschuhHwa01b}, 
or $\phi(N)=1-N^*/N+c(N^*/N)^2$. Both possibilties could not be ruled out completely with the present data, however, the power law leads to better values of $\chi^2$. When the range of $\overline{1-q}$ is restricted to small values ($<0.4$), the finite size corrections $g_h(N)$ can be neglected, and the fit yields values of $\theta\approx 0.35$ as in Ref.~\cite{MarinariPagnaniRicci01}. 

A better understanding of the finite size corrections would thus be needed in
order to determine the precise value of the $\theta$ exponent. At the present stage, taking into account the uncertainty about systematic errors related to the choice of fitting schemes,
our data favour a value of $\theta$ in the range $0.15-0.40$, but we cannot rule
out a scenario
with $\theta=0$.
For models with different pairing energies we obtained similar results, 
favouring a power law for the scaling of excitation energies, with 
exponents in the above-mentioned range of uncertainty.

\section{$\epsilon$-coupling at finite temperature}

Coupling two copies of the system can also be used in its original form at finite temperature\cite{CaraccioloParisi90a,Mezard90}. We consider two RNA 
strands with the same random sequence coupled by the Hamiltonian
\begin{equation}
	H_{\rm tot}({\mathcal S}_1,{\mathcal S}_2;\epsilon)=H({\mathcal S}_1)+H({\mathcal S}_2)+\epsilon q({\mathcal S}_1,{\mathcal S}_2).
\label{Htot}
\end{equation}
The thermal average of the overlap as a function of $\epsilon$ is given by
\begin{equation}
	\langle q\rangle(\epsilon) = \frac{\sum_{{\mathcal S}_1,{\mathcal S}_2}q({\mathcal S}_1,{\mathcal S}_2)\exp(-\beta H_{\rm tot}({\mathcal S}_1,{\mathcal S}_2;\epsilon))}{\sum_{{\mathcal S}_1,{\mathcal S}_2}\exp(-\beta H_{\rm tot}({\mathcal S}_1,{\mathcal S}_2;\epsilon))} = \frac{\langle q({\mathcal S}_1,{\mathcal S}_2)\exp(-\beta \epsilon q({\mathcal S}_1,{\mathcal S}_2))\rangle_{1,2}}{\langle \exp(-\beta \epsilon q({\mathcal S}_1,{\mathcal S}_2))\rangle_{1,2}}.
\label{qmean}
\end{equation}
Here, $\langle\rangle$ denotes the average over the Gibbsian ensemble of the coupled two replica system while $\langle\rangle_{1,2}$ denotes the average with independent Boltzmann weights for the two replicas. Note that the partition function corresponding to the coupled Hamiltonian (\ref{Htot}) is simply related to the Laplace transform of $P(q)$ with respect to $\beta \epsilon$. 
We can rewrite Eq.~(\ref{qmean}) as
\begin{equation}
	\langle q\rangle(\epsilon) = -\frac{1}{\beta}\frac{\partial}{\partial\epsilon}\ln
\left\{\left\langle \sum_{{\mathcal S}_2} \exp[-\beta H({\mathcal S}_2)-\beta\epsilon q({\mathcal S}_1,{\mathcal S}_2)] \right\rangle_1 \right\}
\equiv -\frac{1}{\beta}\frac{\partial}{\partial\epsilon}\ln\left[\langle Z_2(\epsilon;{\mathcal S}_1)\rangle_1\right].
\label{qmean2}
\end{equation}
where the average $\langle\rangle_{1}$ extends only over the first replica.
The averages over the secondary structure of a single RNA-strand are calculated by sampling the Gibbs ensemble as in\cite{Higgs96}: The partition functions $Z_{i,j}$ corresponding to connected substrands $\{i,j\}$ are calculated recursively and can then be used to generate secondary structures with their corresponding Boltzmann weight.  
Since the coupling is not expected to alter the single sequence statistics significantly, we approximated the righthand side of Eq.~(\ref{qmean2}) by calculating an average over 20 randomly sampled structures ${\mathcal S}_1$ for each of which $Z_2(\epsilon;{\mathcal S}_1)$ is calculated exactly. We verified that the results of more extensive samplings lie within the error bars of the disorder average.

In Fig.~\ref{Fig_scaling} we plot the data obtained for $T=0.05$ and $T=0.25$.
The data collapse works best with a free energy exponent $\theta=0.2-0.3$ at $T=0.05$ which coincides with the zero temperature exponent. For $T=0.25$ we have to choose $\theta=1$ to superpose all the curves. This is the same exponent that one finds in the case of homogeneous RNA as is shown in the next paragraph. 
 $\theta=1$ can no longer be interpreted as a free energy exponent at high temperatures. It can be understood in terms of a flat free energy landscape and uncorrelated behaviour of individual bases. A study of the homogeneous case is instructive
in this respect.

\section{$\epsilon$-coupling in the homogeneous case}

Let us consider the problem of RNA with homogeneous base pairing energy $e_{i,j}\equiv e$. In this case it is possible to calculate analytically the asymptotic ($N\rightarrow\infty)$ limit of the partition function of two coupled replicas with Hamiltonian (\ref{Htot}) in close analogy to~\cite{BundschuhHwa01b}: For each configuration of the two replicas we determine the common bonds and the bases which are unpaired in both sequences. The contribution of these common elements is split into a connected part that vanishes for $\epsilon=0$, and a disconnected part corresponding to uncoupled strands. Thus, the contribution of a configuration with $n$ common elements (bonds or unpaired bases) is split into a sum of $2^n$ terms. In each such term we determine the $m$ (connected) common bonds and the $f$ common unpaired bases which are not embraced by a connected common bond. Summing over all possibilities to arrange those $m+f$ connected common elements on the RNA strand we arrive at a recursion relation for the total partition function which reads
\begin{equation}
	\label{ztot}
	Z_{\rm tot}(N)=\sum_{f\geq 0}\sum_{m\geq 0}\sum_{l_1\dots l_m\geq 2} \left( \begin{array}{c} m+N-\sum l_i\\m,f\end{array}\right) Z_1^2(N-\sum_{i=1}^m l_i-f)g_u^f\prod_{i=1}^m \left[g_dZ_{\rm tot}(l_i-2)\right],
\end{equation}
where $g_u=e^{-\beta\epsilon/N}-1$ and $g_d=e^{-2\beta e}(e^{-2\beta\epsilon/N}-1)$ are the connected couplings. The interior of each connected common bond of length $l_i$ is treated as a two replica system with $N$ replaced by $l_i-2$. Note that here we use $s=1$ for the smallest permissible loop. $Z_1(N)$ is the partition function of one replica with $N$ bases. The factor $Z_1^2(N-\sum l_i-f)$ arises from all bonds that can be distributed on the remaining free part of the strands outside the $m$ connected common bonds, excluding the $f$ connected unpaired bases. The combinatorial factor counts the possibilities to align $m$ connected common bonds, $f$ unpaired common bases and $N-\sum l_i-f$ free bases. Of course, we have to require $\sum l_i+f\leq N$.

Let us fix the values of $g_u$ and $g_d$ for a moment and introduce the generating function of $Z_{\rm tot}$ for which we can derive a recursion relation using (\ref{ztot}), 
\begin{equation}
	\Xi(\zeta)=\sum_{N=0}^\infty Z_{\rm tot}(N)\zeta^N=\frac{1}{1-g_d\zeta^2\Xi(\zeta)-g_u\zeta}\Xi_2\left(\frac{\zeta}{1-g_d\zeta^2\Xi(\zeta)-g_u\zeta}\right).
	\label{Xi}
\end{equation}
Here, $\Xi_2(\zeta)=\sum_{N=0}^\infty Z_1^2(N)\zeta^N$ denotes the generating function of $Z_1^2(N)$.
$Z_1(N)$ satisfies the recursion relation,
\begin{equation}
	Z_1(N)=Z_1(N-1)+\sum_{k=0}^{N-2}g_0Z_1(k) Z_1(N-k-2),
\end{equation}
with $g_0=e^{-\beta e}$, and the corresponding generating function $\Xi_1(\zeta)$ can be obtained explicitly,
\begin{equation}
	\Xi_1(\zeta)=\frac{1-\zeta-\sqrt{(1-\zeta)^2-4g_0\zeta^2}}{2\zeta^2g_0}.
\end{equation}
The behaviour of $Z_1(N)$ for large $N$ can be derived from this expression by inverse Laplace transform, see \cite{DeGennes68, Waterman78, BundschuhHwa01b}. Asymptotically, one finds
$	Z_1(N)=c\zeta_1^{-N}/N^{3/2}$,
where $\zeta_1=(1+2\sqrt{g_0})^{-1}$ is the smallest singularity of $\Xi_1(\zeta)$.
Similarly, the smallest singularity $\zeta_*$ of $\Xi(\zeta)$ determines the leading behaviour of $Z_{\rm tot}$. One can check that, since $g_d$ is negative, $\zeta_*$ is always determined by the singularity of $\Xi_2$ on the RHS of Eq.~(\ref{Xi}), i.e., $\zeta_*/[1-g_d\zeta_*^2\Xi(\zeta_*)-g_u\zeta_*]=\zeta_1$, or explicitly,
\begin{equation}
	\zeta_*=\zeta_1/[1+g_d\zeta_1^2\Xi(\zeta_1)+g_u\zeta_1].
\end{equation}
Asymptotically, the two replica partition function behaves as $Z_{\rm tot}(N)={\tilde c}\frac{\zeta_*^{-N}}{N^{3/2}}$.
We can now calculate $\langle q\rangle(\epsilon)$ by taking the logarithmic derivative of $Z_{\rm tot}(N)$ recalling the dependence of $g_u$ and $g_d$ on $\epsilon/N$,
\begin{equation}
	\label{qhom}
	\langle q\rangle(\epsilon,N) = -\frac{\partial\ln Z_{\rm tot}(N)}{\beta\partial\epsilon}\approx \frac{N}{\beta}\frac{\partial\ln \zeta_*}{\partial\epsilon}
 = \frac{2g_0^2\zeta_1^2\Xi_2(\zeta_1)e^{-2\beta\epsilon/N}+\zeta_1e^{-\beta\epsilon/N}}{1+g_d\zeta_1^2\Xi_2(\zeta_1)+g_u\zeta_1}.
\end{equation}
From Eq.~(\ref{qhom}) it is clear that $\langle q\rangle =q(\epsilon/N)$ and thus $\theta=1$ in the homopolymer. 
We can understand this scaling behaviour as a lack of long-range correlations in the system (A simple example of such a behaviour  is the case of $N$ 
uncoupled Ising spins in a random field under $\epsilon$-coupling). 
In order to affect the overlap significantly one has to introduce an extensive perturbation which corresponds to a finite force acting on each base and typically leads to a large number of independent local rearrangements.

\section{Phase transition}
We will now use the scaling exponent $\theta$ to define an order parameter for the presumed phase transition in the disordered system. Consider the effect of a small extensive coupling between two replicas, i.e., $\epsilon=\delta N$. In a phase with $\theta<1$ where typical excitation energies are subextensive ($o(N)$), the coupling always dominates in the thermodynamic limit the replicas will have no overlap ($q=0$) for a repulsive coupling $\delta>0$ while they are locked together ($q=1$) when the coupling is attractive ($\delta<0$).
However, if $\theta=1$ an extensive coupling is a marginal perturbation. In the homogeneous case we can directly check from Eq.~(\ref{qhom}) that small perturbations have a negligible effect, and $\langle q\rangle$ is continuous as $\delta \to 0\pm$. From the numerical data in Fig.~\ref{Fig_scaling} we see that the same is true in the high temperature phase of disordered sequences which leads us to define the order parameter
\begin{equation}
	\label{OP}
	\phi=\lim_{\delta\to 0}\left\{\lim_{N\to \infty}\left[\overline{q}(\epsilon=-\delta N;N)-\overline{q}(\epsilon=\delta N;N)\right] \right\}. 
\end{equation} 
For a homopolymer $\phi\equiv 0$ at all temperatures. For disordered sequences $\phi$ jumps discontinuously at the critical temperature from $0$ in the high temperature phase to $q_{\rm max}-q_{\rm min}=1$ in the glassy phase. This provides a precise mathematical definition of the transition temperature.     

\section{Conclusion}

	We have studied a simple model for the 
folding of RNA, taking as a probe the susceptibility to
the introduction of a repulsion between two identical clones of the system.
 Our method clearly distinguishes a low temperature glassy
regime, identical to the zero temperature case, where the  excitation energies
scale with a power law of the length, with a small  
scaling exponent. On the contrary the 
 high temperature liquid-like phase is found to behave similarly to a homopolymer, both having a relatively flat free energy landscape and no long-range correlations in the base pairing pattern. The result for the exponent $\theta$ gives way to the definition of an order parameter which jumps discontinuously at the critical temperature.

\acknowledgments

It is a pleasure to thank R. Bundschuh, T. Hwa, M. L{\"a}ssig, E. Marinari, O.C. Martin, G. Parisi, L. Peliti
and F. Zuliani for helpful discussions. 
F. Krzakala and M. M\"uller acknowledge 
a fellowship from the MRT. The LPTMS is an Unit\'e de Recherche de
l'Universit\'e Paris~XI associ\'ee au CNRS.

\bibliographystyle{prsty}
\bibliography{../Bib/references}

\end{document}